\documentclass[twocolumn]{aastex62}
\pdfoutput=1 

\usepackage{newtxtext,newtxmath}

\usepackage{epsfig}
\usepackage{amsmath,amstext}
\usepackage[T1]{fontenc}
\usepackage{footmisc}
\usepackage{natbib}
\usepackage[figure,figure*]{hypcap}

\newcommand{\Sp}{{\it Spitzer\/}}
\newcommand{\Her}{{\it Herschel\/}}
\usepackage{graphicx}	
\usepackage{amssymb}	

\begin{document}

\title{Massive Young Stellar Objects and Outflow in 
the Infrared-Dark Cloud G79.3+0.3}

\author{Anna S.E. Laws}
\affiliation{Astrophysics Group, University of Exeter, 
Stocker Road, Exeter, EX4 4QL, U.K.}
\affiliation{School of Physics and Astronomy, University of Southampton, 
Highfield, Southampton SO17 1BJ, UK}
\author{Joseph L. Hora}
\affiliation{Center for Astrophysics | Harvard \& Smithsonian,
60 Garden Street, Cambridge, MA 02138, USA
}
\author{Qizhou Zhang}
\affiliation{Center for Astrophysics | Harvard \& Smithsonian,
60 Garden Street, Cambridge, MA 02138, USA
}
\correspondingauthor{Anna S.E. Laws}
\email{al630@exeter.ac.uk}

\begin{abstract}
G79.3+0.3 is an infrared-dark cloud (IRDC) in the Cygnus-X complex that is home to  
massive deeply-embedded Young Stellar Objects (YSOs).
We have produced a Submillimeter Array (SMA) 1.3~mm continuum image and $^{12}$CO line maps of the eastern section of 
G79.3+0.3 in which we detect five separate YSOs.
We have estimated physical parameters for these five YSOs  
and others in the vicinity of G79.3+0.3 by fitting existing 
photometry from \Sp, \Her, and ground-based telescopes to 
spectral energy distribution (SED) models.
Through these model fits we find that the most massive YSOs seen in
the SMA 1.3~mm continuum emission have masses in the $5 - 6$~$M_{\sun}$ range.
One of the SMA sources was observed to power a massive  collimated $^{12}$CO outflow
extending at least 0.94~pc in both directions from the protostar, with a total mass of 0.83~$M_{\sun}$ and a dynamical timescale of 23~kyr.
\end{abstract}




\section{Introduction}

\begin{figure*}
    \centering
    \includegraphics[width=1.8\columnwidth]
                    {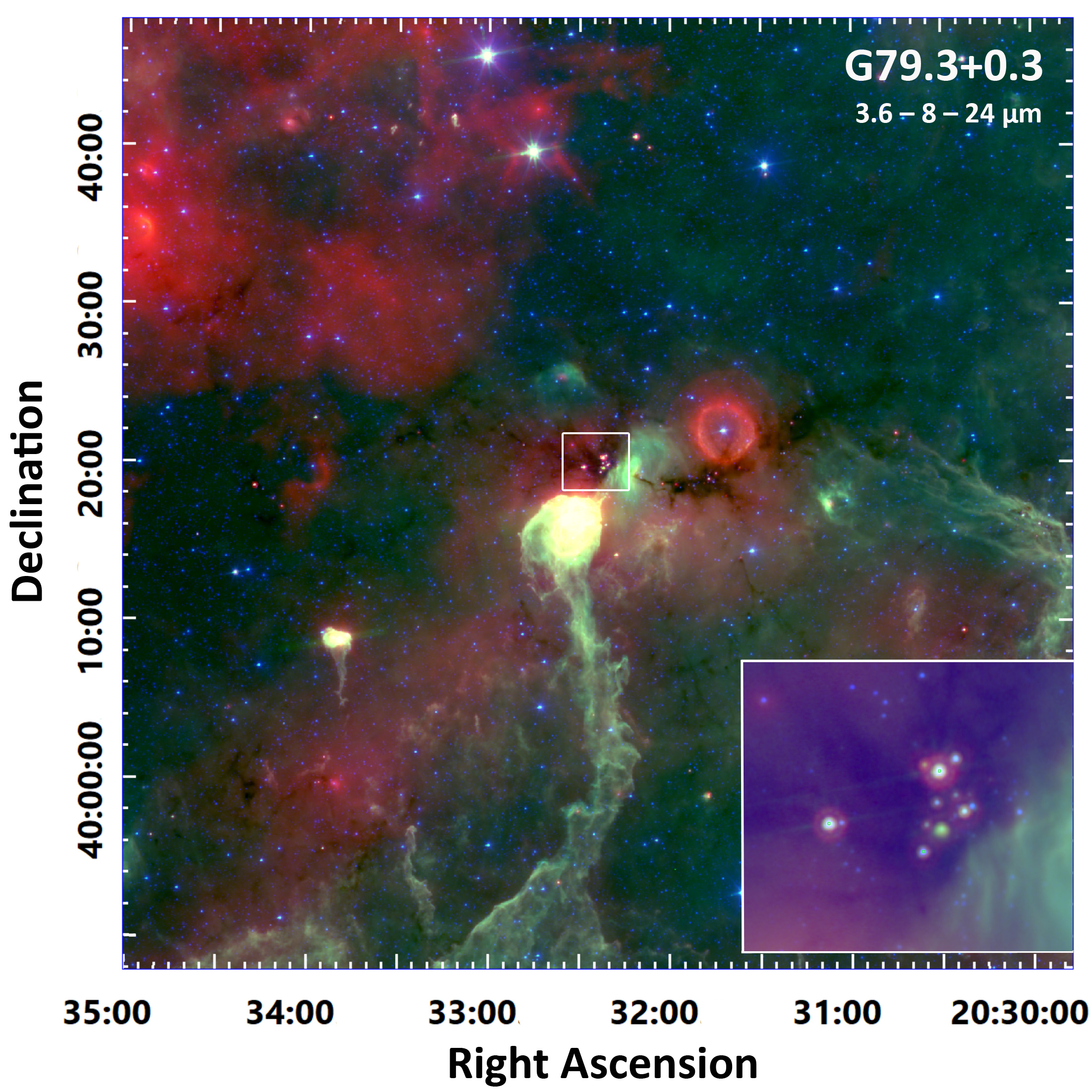}
    \caption{A near-IR view of the G79.3+0.3 and its surroundings.
             The region in G79.3+0.3 observed
             with the SMA is directly in 
             the center of this image, highlighted with a white square 
             that is enlarged within the inset. 
             The inset shows the more immediate surroundings of the 
             region of the IRDC imaged with the SMA, spanning about 3.7$\times$3.0 arcmin or 1.5$\times$1.2~pc. The IRDC
             can be seen as the dark area immediately surrounding the 
             central YSOs, continuing to the top and left sides of the 
             inset.
             These are composite images using three wavebands,  
             3.6~\micron\ (blue), 8~\micron\ (green), 
             and 24~\micron\ (red) from the 
             \Sp\ Space Telescope.}
    \label{fig:spitzer}
\end{figure*}

Massive stars are known to form in clusters in dense molecular clouds, but there are several aspects of the formation process that are not well-understood. For example, it is not yet known which is the dominant massive star formation mechanism \citep[i.e., monolithic collapse, competitive accretion, and stellar collisions and mergers; e.g. see reviews by][]{zinnecker2007, mckee2007, tan2014, motte2017}. In order to determine the critical processes involved, it is necessary to find and study examples of massive stars in various stages of formation. This task is made difficult by the relatively short formation period and lifetime of massive stars, the low numbers of massive stars given by the mass function of stellar clusters, and the fact that the birthplace of these stars are in the cores of dense molecular clouds, making the massive Young Stellar Objects (YSOs) difficult to observe. 

With the availability of sensitive infrared surveys from the \Sp\ Space Telescope \citep{werner2004}, and high-resolution (sub)millimeter interferometers such as the Submillimeter Array (SMA), it is possible to probe these deeply-embedded YSOs near the beginning of their formation process. Several investigations have identified and studied the objects known as infrared dark clouds (IRDCs), which have been often found to host sites of massive star formation 
\citep[e.g.][]{wang2006,rathborne2006,cyganowski2008,peretto2010}. The SMA has been used to study young massive embedded clusters, revealing information on the earliest stages of formation and mass accretion \citep[e.g.,][]{zhang2009,zhang2011, wang2013,wang2014}. By combining the IR and submm data, we can study the formation of YSOs and the cluster from its earliest stages onward.

Cygnus-X is one of the most massive molecular cloud complexes in the 
Galaxy with a total mass of 3x$10^{6} M_{\sun}$ 
\citep{schneider2006}, and is one of the nearest 
massive star-forming complexes at a distance of $1.40\pm0.08$ kpc \citep{rygl2012}.
Along with YSOs, Cygnus-X includes 800 distinct \ion{H}{2} regions (evidence of 
widespread massive star formation),
several OB associations, and several Wolf-Rayet and O3 stars 
\citep{beerer2010}. The Cygnus OB2 star cluster 
contains hundreds of massive OB and O stars, with a stellar mass of $\sim 1.7\times10^4M_{\sun}$ \citep{wright2015}.
Altogether this makes the complex an excellent target for 
observing massive stars and their formation.

In this paper, we study a region in the IRDC G79.3+0.3 \citep[][see Figure~\ref{fig:spitzer}]{carey1998,redman2003}. This IRDC is one of the largest in the Cygnus-X region, and lies in the central subsection of Cygnus-X near DR15, an \ion{H}{2} region that has been previously targeted as a candidate for high-mass star formation \citep{riveragalvez2015,schneider2016}. 
Star formation is especially promising in DR15 with its molecular pillar
and prominent bright envelope (as seen in Figure~\ref{fig:spitzer}), 
though the IRDC is likely not interacting with the pillar 
\citep{riveragalvez2015}. Located close to the IRDC is the luminous blue variable (LBV) candidate star G79.29+0.46, seen as the blue star in the middle of the circular red nebula to the right of center in Figure~\ref{fig:spitzer}. Studies by \citet{umana11} with the EVLA and observations of NH$_3$ (1, 1) and (2, 2) emission by \citet{rizzo2014} showed evidence that the LBV is interacting with the IRDC. 
Following \citet{rizzo2014}, we assume that the IRDC G79.3+0.3 is located at the same distance as Cygnus-X at 1.4 kpc. 

A total of 226 young stellar sources have previously been seen in the 
DR15 area, and the Class 0 and I objects are mostly intermediate- to 
high-mass stars \citep{riveragalvez2015}.
G79.3+0.3 has a molecular mass of 803~$M_{\sun}$, which is average for the sample 
of 45 massive star-forming IRDCs given in \citet{ragan2012}, but high  
compared with the other Cygnus-X IRDCs (\citeauthor{calahan2017}, in prep.)
thus making the cloud a likely location of massive YSOs in Cygnus-X.
G79.3+0.3 is much closer than most IRDCs, at a distance of 1.4 kpc 
compared with the mean distance to classical IRDCs of $>$3 kpc 
\citep{ragan2012}, giving a greatly improved chance of resolving
individual objects within the cloud. 

YSOs can produce molecular outflows as a consequence of the accretion 
process. As protostars gain mass through accretion, infalling gas with excess angular momentum is ejected in a wind \citep{shu2000,shang2007} that accelerates the ambient material to higher velocities seen as molecular outflows in CO and SiO \citep[e.g.][]{lee2002,palau2006,zhang1995b}. Surveys of massive star forming regions \citep{zhang2001,zhang2005,beuther2002b} revealed ubiquitous outflows associated with massive protostars. The mass, momentum and energy are typically orders of magnitude larger than those of the low mass stars. These studies provided crucial observational constraints on the formation process of massive stars. In more recent years, outflows and H$_2$O masers were also detected in massive IRDC clumps \citep{wang2006,wang2012,wang2014}, indicative of star forming activities in these clouds. Thanks to the physical connections with accretion, outflow energetics provide valuable insights into the accretion process of protostars.  

In this paper we present the results of our study of
the YSOs in G79.3+0.3, including new high-resolution SMA interferometric
observations of a portion of the
G79.3+0.3 IRDC. In Section~\ref{sec:data} we provide the details of the
new observations, and also describe the photometry we used
from the \Sp\ and \Her\ missions and from ground-based
near-IR surveys. Section~\ref{sec:results} gives the results of our SED
modelling of YSOs. The analysis of the outflow detected in
SMA source 1 is described in Section~\ref{sec:discussion}, along with discussion of former studies of these objects. We summarize the main findings in Section~\ref{sec:summary}.

\section{Observations and Data Reduction}
\label{sec:data}
\subsection{SMA}
The Submillimeter Array\footnote{The Submillimeter Array is a joint project between the Smithsonian 
Astrophysical Observatory and the Academia Sinica Institute of Astronomy 
and Astrophysics and is funded by the Smithsonian Institution and the 
Academia Sinica.} \citep{ho2004} is an interferometer consisting of 
eight 6~m-diameter antennas operating at millimeter and sub-millimeter wavelengths 
located near the summit of Maunakea, Hawaii. 
The SMA observations of G79.3+0.3 were carried out on July 8th 2012 and August 
3rd 2012 using the Compact and Subcompact array configurations 
respectively. The array used the 230 GHz receivers tuned to an LO frequency of 224.92 GHz. With an IF frequency of 4 to 8 GHz the observations covered sky frequencies of $216.9 - 220.9$ GHz in the lower sideband (LSB) and $228.9 - 232.9$ GHz in the upper sideband (USB). The digital correlator was configured to a uniform spectral resolution of 0.8~MHz per spectral channel across the entire 8~GHz band.  The FWHM of the primary beam response of the 6~m antennas is approximately $55''$. The observations employed a total of 10 pointings separated by half of the FWHM of the primary beam. We used MWC349A and QSO 2007+404 as the time dependent gain calibrators. The spectral bandpass was calibrated using 3C279. The flux calibration was carried out through observations of a known flux source Titan. The uncertainty of the absolute flux calibration is about 15\%. The system temperatures during the observations were from 120 to 160 K. 

We calibrated the data using the MIR software (Millimeter Interferometer
Reduction)\footnote{https://www.cfa.harvard.edu/$\sim$cqi/mircook.html} following the procedure outlined at the SMA data reduction website.
The calibrations include systemic temperature correction,  bandpass, time dependent 
gain variations, and flux calibration.
The calibrated visibilities were then exported to MIRIAD and CASA for imaging.
We separated the continuum and line emission in visibilities and then Fourier transformed and cleaned.  We used $tclean$ task in CASA with Briggs weighting and a robust parameter of 0.5. The resulting 1.3~mm continuum image has a 1$\sigma$ rms noise level of 4 mJy~beam$^{-1}$. The 1$\sigma$ rms in the 1 km~s$^{-1}$ channels is 140 mJy~beam$^{-1}$. The spatial resolution of the continuum image is approximately 2\farcs5 FWHM.

\subsection{Infrared Photometry}

We used four photometry catalogs in our study of
Cygnus-X. The first contained photometry from the following infrared 
surveys: 2MASS \citep{2mass2006}, UKIDSS \citep{ukidss2006}, and \Sp\
\citep{spitzercat2011}. The catalogs contain over three million objects in the 
region, of which we classified 30,902 as YSOs (Classes 0, I and II) using the methods 
described by \citet{gutermuth08}. There are 
28 objects classified as YSOs in the field observed by the SMA. 
For the J, H, and K bands, we used data from the two ground-based 
surveys UKIDSS and 2MASS. The UKIDSS survey has better resolution than 2MASS 
($\sim$1\arcsec\ vs. 4\arcsec, respectively) but is saturated for objects 
brighter than 11~mag, so we used 2MASS data for recorded magnitudes below 
11~mag and UKIDSS data otherwise. The IRAC images have FWHM resolutions of 1\farcs66, 1\farcs72, 1\farcs88, and 1\farcs98 for the 3.6, 4.5, 5.8, and 8~\micron\ bands, respectively \citep{fazio04}. The MIPS instrument has a 
24~\micron\ resolution of 6\arcsec \citep{rieke04}.

We utilized \Her\ data from three separate catalogs, one
containing  photometry for 579 YSOs in Cygnus-X detected by 
both \Sp\  and \Her/SPIRE at 250, 350, and 500~\micron 
~\citep{kirk2014}.
We also downloaded data from two catalogs containing photometry for 698 YSOs in Cygnus-X
observed with the PACS 70 and 160~\micron\ bands \citep{pacscatalogs2017}. The spatial resolution of \Her\ is approximately 5\arcsec, 13\arcsec, 18\arcsec, 25\arcsec, and 36\arcsec\ for the 70, 160, 250, 350, and 500~\micron\ bands, respectively.
%

\begin{figure*}
    \centering
    \includegraphics[width=0.8\textwidth]{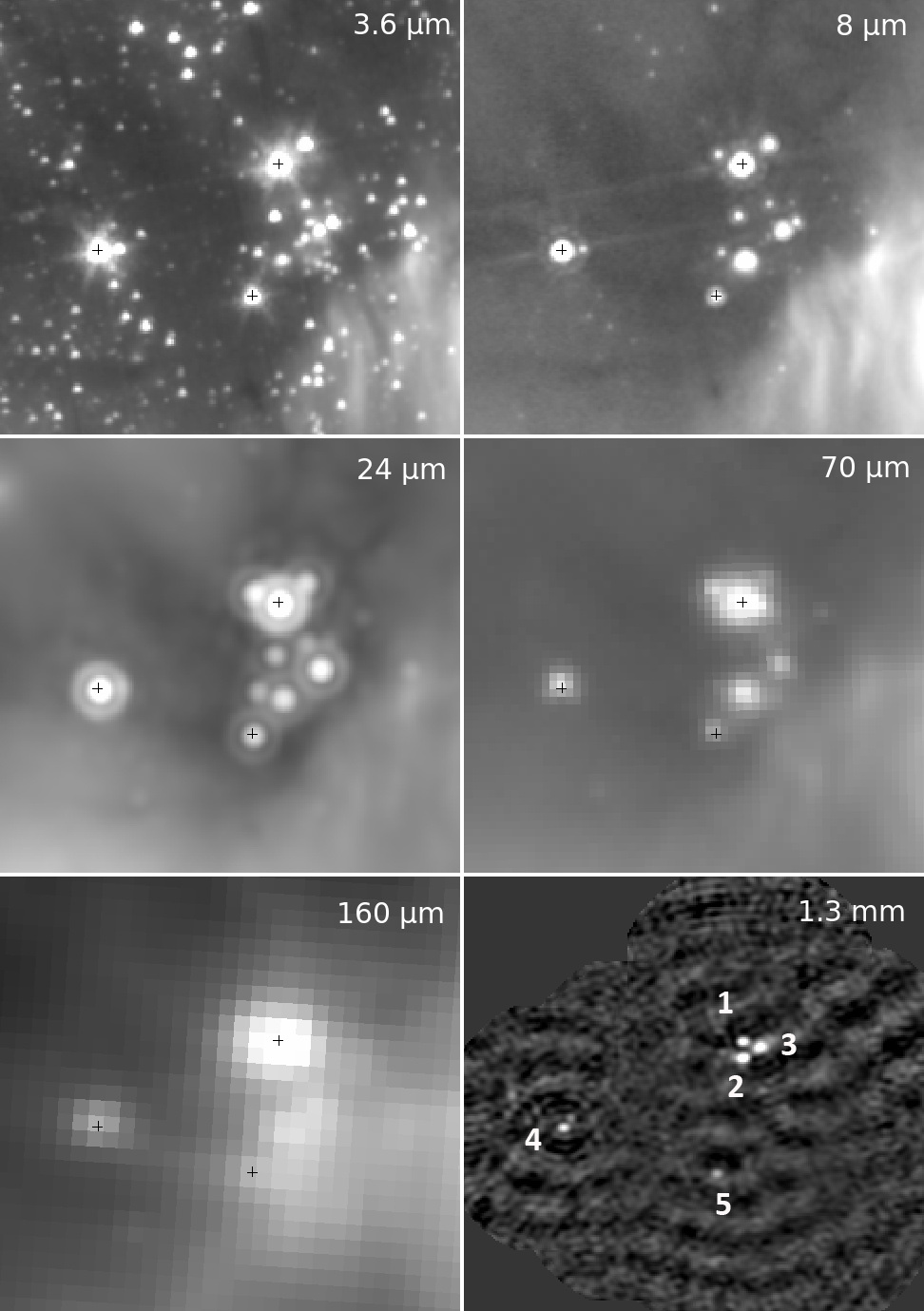}
    \caption{Views of G79.3+0.3 as seen in various infrared wavelengths
            and by the SMA. The same 3.4\arcmin$\times$3\arcmin\ region is shown in each panel. In the lower right panel the SMA image is shown with each of the continuum sources labeled. The positions of these sources are given in Table \ref{tab:sma_flux}. The SMA objects 1, 
            4, and 5 (marked with the crosses in the five other panels) are each clearly visible in every image up to and 
            including the PACS 
            160~\micron image. The SMA sources 2 and 3 are not detected at wavelengths shorter than 70~\micron. At 70 and 160~\micron, evidence for SMA sources 2 and 3 is present but they are not fully resolved because of the lower spatial resolution at these wavelengths.
            }
    \label{fig:dr15_panels}
\end{figure*}


\section{Results}
\label{sec:results}
\subsection{SMA Continuum}
A comparison of G79.3+0.3 at different wavelengths is shown in 
Figure~\ref{fig:dr15_panels}, with the new SMA 1.3~mm continuum image shown in the lower right panel.  
Five distinct continuum peaks (labeled SMA objects 1 
through 5) can be seen, and each object's position and flux is given in 
Table~\ref{tab:sma_flux}. The position of SMA sources 1, 4, and 5 are indicated by crosses overlaid on the \Sp\ and \Her\ images in Figure~\ref{fig:dr15_panels}.
Two of these objects (labeled 2 and 3) show extended emission coinciding with molecular outflows. We will explore outflows
identified in the CO emission in Section~\ref{sec:results_12co}.

\begin{deluxetable}{crrrr}
\tablecaption{Fluxes and masses for the 1.3 mm sources\label{tab:sma_flux}}
\tablecolumns{5}
\tablewidth{0pt}
\tablehead{
\colhead{SMA} & \colhead{R.A.}        & \colhead{Dec.} & 
\colhead{Flux density}  & \colhead{Mass}\\
\colhead{ID} & \multicolumn{2}{c}{(J2000.0)}    & \colhead{(mJy)}& \colhead{(M$_{\sun}$)}
  }
\startdata 
   1   & 20:32:22.1	&	40:20:17.1 & $65.7 \pm 1.2$ 
                                  & $2.36 $\\
  2   & 20:32:22.0	&	40:20:09.7 & $128 \pm 2\phantom{.0}$
                                  & $4.60 $ \\
  3   & 20:32:21.4  &	40:20:14.1 & $150 \pm 1\phantom{.0}$
                                  & $5.39 $\\
  4   & 20:32:28.6  &   40:19:41.6 & $61.6 \pm 0.8$ & $2.21 $\\
  5   & 20:32:23.0	&	40:19:22.7 & $39.0 \pm 0.4$ & $1.40 $\\
  \enddata
\end{deluxetable}

The continuum flux measurements can be used to calculate the mass of the gas
surrounding the objects following the method in \citet[eqn.1]
{motte2007}. We assumed values of dust emissivity $\kappa_{1.3 \rm{mm}} = 0.01$~cm$^2$~g$^{-1}$, $T_{\rm dust} = 15$K, and a gas-to-dust mass ratio of 100:1. The masses are given in Table~\ref{tab:sma_flux}. Following Motte et al., we note that the mass estimates are uncertain by
a factor of 2 due to uncertain dust emissivity. The individual
masses can vary by $\pm$30\% relative to each other when dust temperatures
vary from 15 to 25 K, or by $\pm$50\% for the 10 to 20 K
temperature range.

Using the assumed distance to G79.3+0.3 of $1.4\pm0.08$ kpc, 
the largest project separation between SMA objects is 0.76 pc between objects 3 
and 4, and the smallest separation is 0.069 pc between 1 and 3. Objects 1, 
4 and 5 form a triangle with roughly equal side lengths. The separations 
are: 1--4, 0.70 pc; 4--5, 0.58 pc; and 1--5, 0.38 pc, all with uncertainties 
on the order of $\pm$10\%.
%

\subsection{SED Modeling}
We used SEDFitter\footnote{https://github.com/astrofrog/sedfitter} v1.0 \citep[][]{robitaille2007} to fit model YSO SEDs to 
each object's photometry and return an estimate of its physical 
parameters, including mass and luminosity. It is necessary to find the 
object parameters by comparing observations with models as direct observational methods cannot distinguish features such as the separate central object, disc, 
and envelope. For objects with multiple model fits, we selected the best 
fit as that with the lowest chi-square value and derived object 
parameters from that fit.
The two constraints fed into the SEDFitter are the distance to the object $1.3\leq\text{d}\leq1.5$ kpc and the range of extinctions $5.0\leq\text{A}_{v}\leq 100$~mag.
We created a custom filter to incorporate the SMA photometry with SEDFitter following 
the documentation \citep{sedfitterdocs}.
The atmospheric transmission at the SMA is 100\% in the frequency ranges 
covered by both upper and lower sidebands \citep{welch1988}, so the filter uses a simple function with 100\% transmittance in these frequency ranges and $0\%$ outside.

\begin{figure*}
   \centering
    \includegraphics[height=0.53\columnwidth]{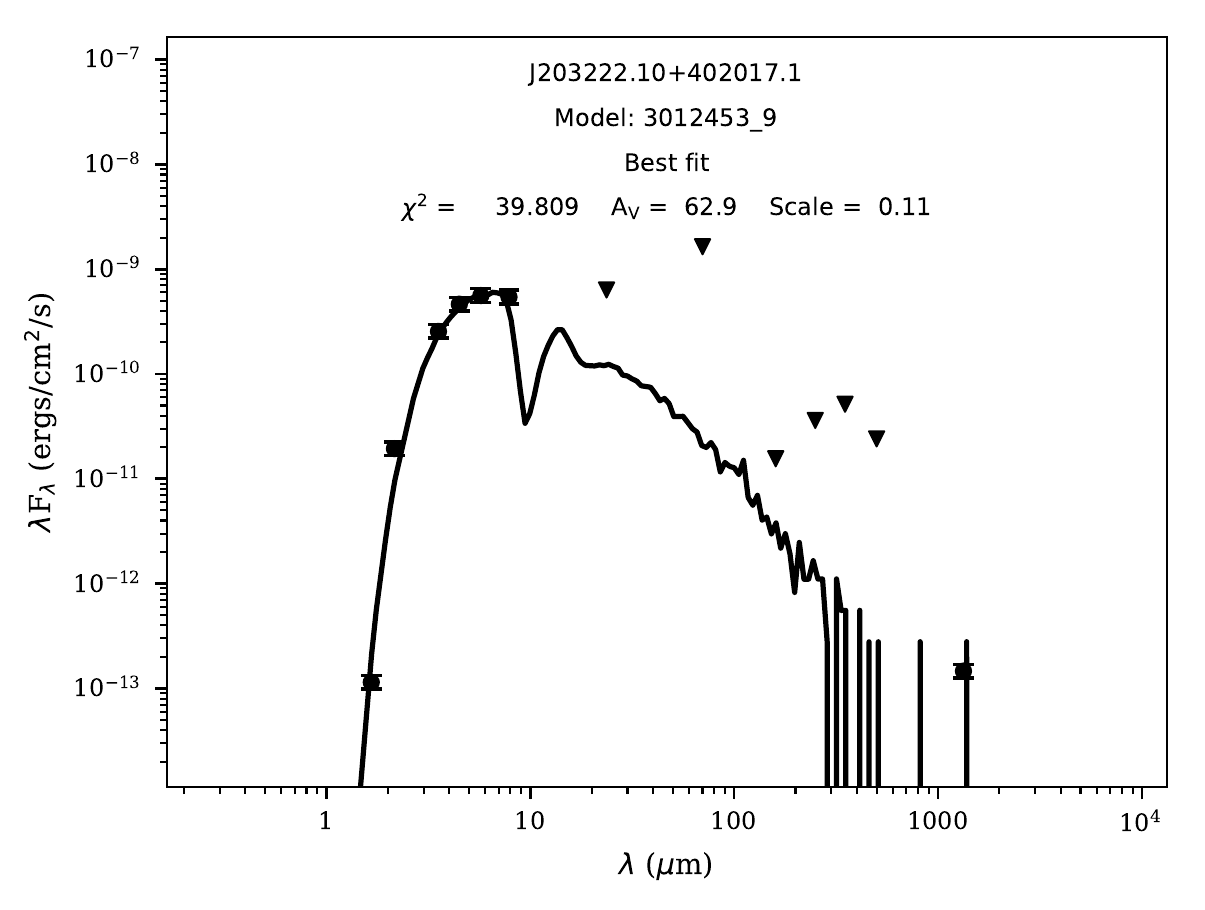}
    \includegraphics[height=0.53\columnwidth]{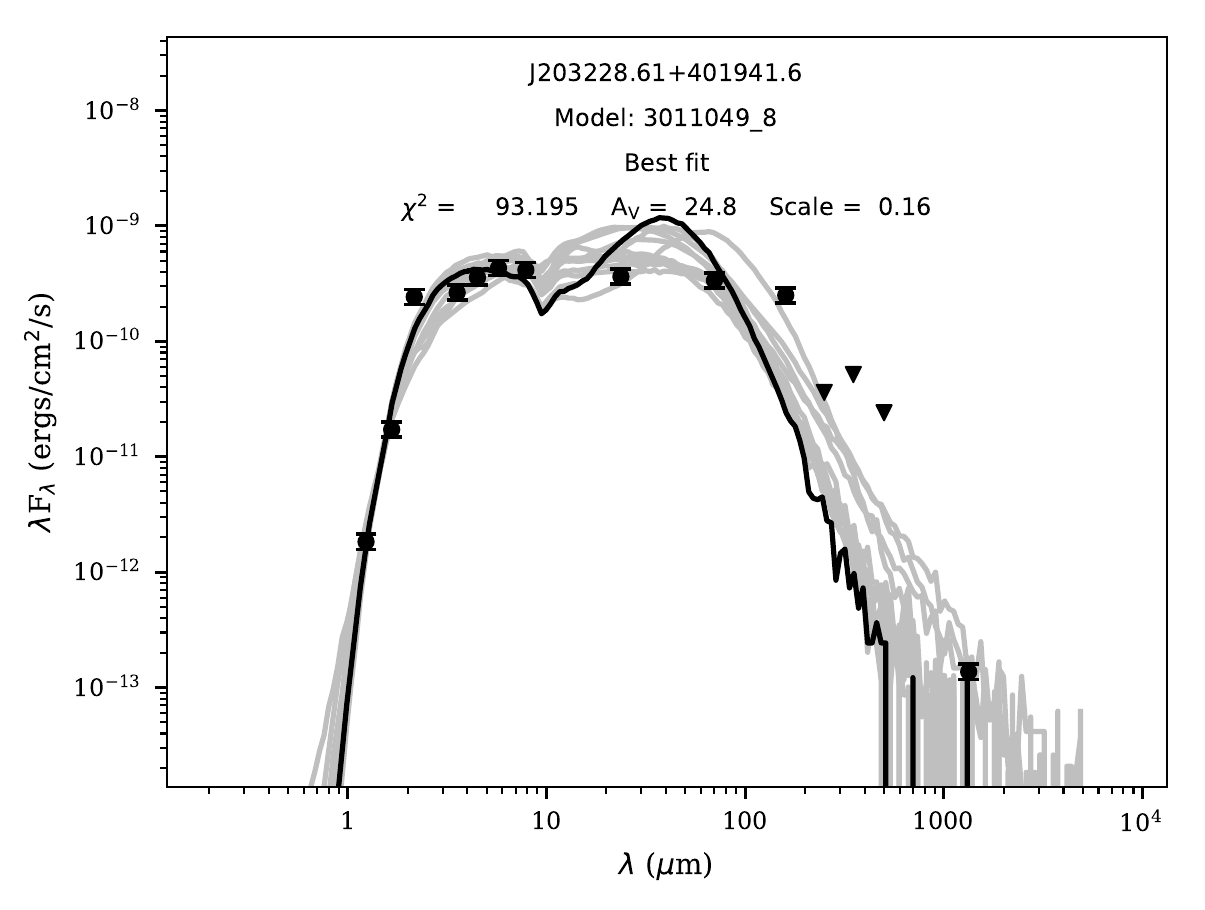}
    \includegraphics[height=0.53\columnwidth]{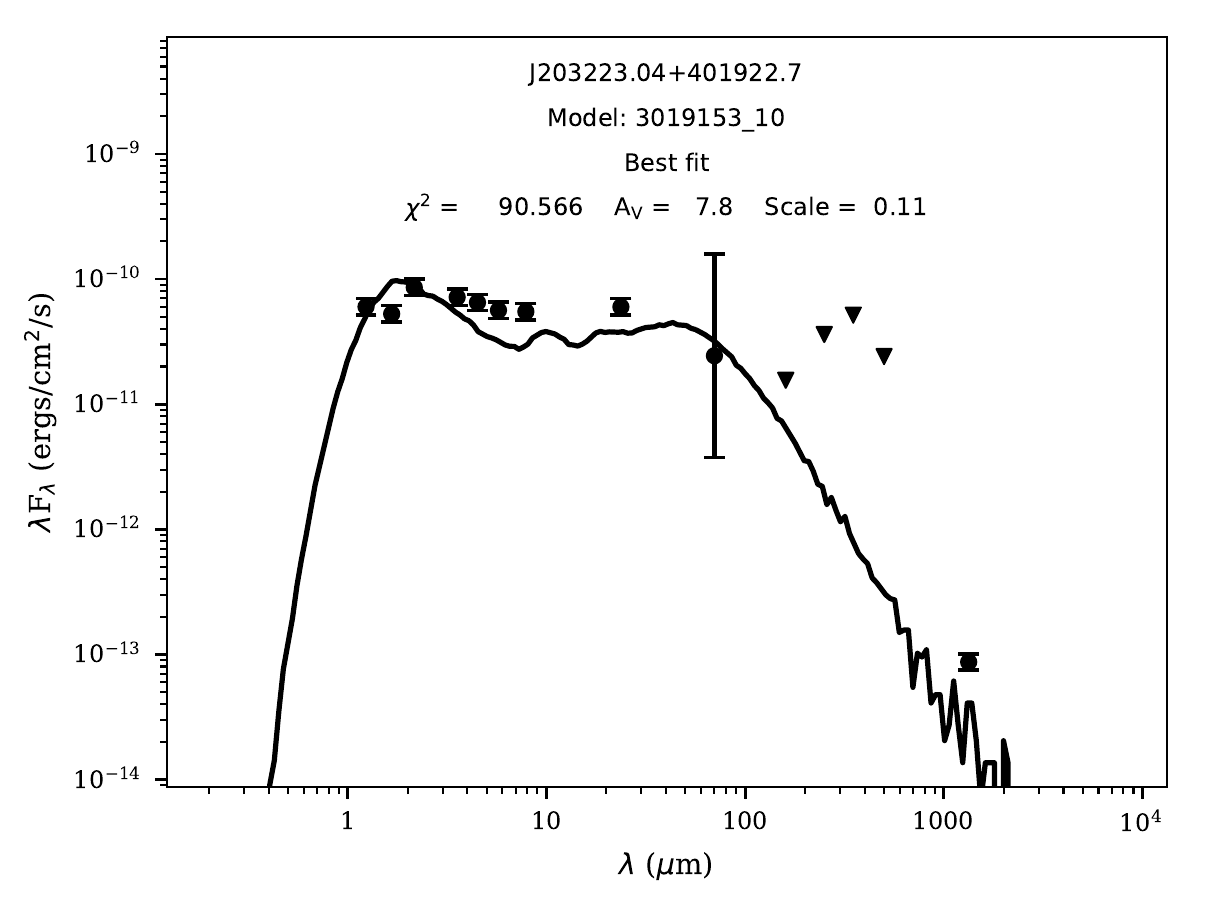}
    \caption{SEDs and the best model fit for SMA objects 1 (left), 4 
             (middle), and 5 (right). Round markers with error bars are
             flux measurements and their uncertainties, and 
             downward-pointing triangles are upper limits for flux. 
             Parameters derived from these SEDs are given in 
             Table~\ref{tab:sed_params}.
             }
    \label{fig:sma_seds}
\end{figure*}

Most YSOs in the \Sp\ catalog have no detection at the longer PACS 
and/or SPIRE wavelengths, constraining these objects to have a flux 
below a certain sensitivity limit at 
these wavelengths. We estimated these limits in Cygnus-X 
using a histogram of the number of objects measured with various fluxes.
The upper limits for undetected sources are as follows: 223 and
827~mJy for the PACS 70 and 160~\micron\ bands respectively; and 3000, 6000, and 4000~mJy for the SPIRE 250, 350, and 500~\micron\ bands respectively.
We similarly estimated a sensitivity limit for the SMA observations by 
assuming that the limit was 10\% of the flux of the faintest distinct 
object. For SMA object 5 at flux 39.0~mJy, this gave a sensitivity limit of 3.90~mJy. 

We used SEDFitter to fit all of the YSOs identified in the Cygnus-X region using color-color and color-magnitude relations \citep{beerer2010}; the fluxes for the objects are given in Table \ref{tab:fluxvalues}. The SEDFitter results for the 28 YSOs within the region observed by the SMA are given in Table \ref{tab:sed_params}. The SED fitting results for the whole of Cygnus-X will be discussed in a later paper.

\startlongtable
\begin{splitdeluxetable*}{rrrccccccccBccccccccccccBccccccccccc}
\tabletypesize{\scriptsize}
\tablecaption{Fluxes for 28 YSOs in the SMA-observed Region of G79.3+0.3\label{tab:fluxvalues} }
\tablewidth{0pt}
\tablehead{
& \colhead{R.A.}& \colhead{Decl.}& & &\multicolumn{4}{c}{Flux Density and Uncertainty (Jy)}\\
\colhead{Name} &\colhead{(J2000.0)} & \colhead{(J2000.0)} & \colhead{{\it J} }& \colhead{{\it J} err} & \colhead{{\it H}} & \colhead{{\it H} err} &\colhead{{\it K}} & \colhead{{\it K} err }& \colhead{3.6} & \colhead{3.6 err} & \colhead{4.5} & \colhead{4.5 err} & \colhead{5.8} & \colhead{5.8 err} & \colhead{8} &\colhead{ 8 err} &\colhead{ 24} & \colhead{24 err} & \colhead{70} & \colhead{70 err} & \colhead{100} & \colhead{100 err} & \colhead{160} & \colhead{160 err} & \colhead{250} & \colhead{250 err} & \colhead{350} & \colhead{350 err} & \colhead{500} & \colhead{500 err} & \colhead{1300} & \colhead{1300 err}
  }
\startdata 
J203222.10+402017.1 & 308.092102 & 40.338074 & \nodata & \nodata & 1.801E+01 & 8.408E-02 & 1.167E+01 & 9.740E-04 & 7.411E+00 & 3.000E-03 & 6.021E+00 & 2.000E-03 & 5.066E+00 & 1.000E-03 & 4.136E+00 & 1.000E-03 & 3.950E-01 & 1.000E-02 & \nodata & 4.431E+02 & \nodata & \nodata & \nodata & \nodata & \nodata & \nodata & \nodata & \nodata & \nodata & \nodata & 6.574E+01 & 1.223E+00 & \\
J203228.61+401941.6 & 308.119202 & 40.328213 & 1.580E+01 & 5.838E-03 & 1.256E+01 & 8.420E-04 & 8.943E+00 & 1.600E-02 & 7.372E+00 & 3.000E-03 & 6.304E+00 & 2.000E-03 & 5.349E+00 & 1.000E-03 & 4.427E+00 & 2.000E-03 & 9.810E-01 & 8.000E-03 & 7.945E+03 & 9.183E+01 & \nodata & \nodata & \nodata & \nodata & \nodata & \nodata & \nodata & \nodata & \nodata & \nodata & 6.160E+01 & 7.900E-01 & \\
J203223.04+401922.7 & 308.096008 & 40.322975 & 1.202E+01 & 6.110E-04 & 1.135E+01 & 4.000E-04 & 1.007E+01 & 1.400E-02 & 8.789E+00 & 2.000E-03 & 8.154E+00 & 2.000E-03 & 7.561E+00 & 3.000E-03 & 6.622E+00 & 5.000E-03 & 2.940E+00 & 3.400E-02 & 3.296E+03 & 6.177E+03 & 6.683E+03 & 8.697E+02 & \nodata & \nodata & \nodata & \nodata & \nodata & \nodata & \nodata & \nodata & 3.903E+01 & 4.480E-01 & \\
J203220.60+401950.1 & 308.085846 & 40.330597 & \nodata & \nodata & \nodata & \nodata & 1.722E+01 & 9.206E-02 & 1.023E+01 & 1.500E-02 & 8.048E+00 & 6.000E-03 & 6.613E+00 & 2.000E-03 & 5.777E+00 & 7.000E-03 & 2.145E+00 & 3.500E-02 & 5.854E+03 & \nodata & \nodata & \nodata & \nodata & \nodata & \nodata & \nodata & \nodata & \nodata & \nodata & \nodata & \nodata & \nodata & \\
J203222.25+401955.8 & 308.092682 & 40.332169 & \nodata & \nodata & 1.483E-01 & 2.225E-02 & 2.121E+00 & 3.181E-01 & 1.447E+01 & 2.171E+00 & 2.760E+01 & 4.140E+00 & 4.499E+01 & 6.748E+00 & 6.267E+01 & 9.400E+00 & \nodata & \nodata & \nodata & \nodata & \nodata & \nodata & \nodata & \nodata & \nodata & \nodata & \nodata & \nodata & \nodata & \nodata & \nodata & \nodata & \\
J203227.86+401942.3 & 308.116089 & 40.328426 & 3.283E-02 & 4.924E-03 & 6.284E-01 & 9.426E-02 & 4.560E+00 & 6.840E-01 & 1.679E+01 & 2.518E+00 & 2.298E+01 & 3.447E+00 & 2.953E+01 & 4.430E+00 & 3.205E+01 & 4.808E+00 & \nodata & \nodata & \nodata & \nodata & \nodata & \nodata & \nodata & \nodata & \nodata & \nodata & \nodata & \nodata & \nodata & \nodata & \nodata & \nodata & \\
J203221.13+402025.6 & 308.088043 & 40.340446 & \nodata & \nodata & 1.310E-01 & 1.964E-02 & 4.769E+00 & 7.153E-01 & 5.447E+01 & 8.170E+00 & 9.893E+01 & 1.484E+01 & 1.515E+02 & 2.272E+01 & 1.872E+02 & 2.808E+01 & 8.494E+02 & 1.274E+02 & \nodata & \nodata & \nodata & \nodata & \nodata & \nodata & \nodata & \nodata & \nodata & \nodata & \nodata & \nodata & \nodata & \nodata & \\
J203220.14+401953.5 & 308.083893 & 40.331524 & \nodata & \nodata & \nodata & \nodata & 7.542E-01 & 1.131E-01 & 2.955E+01 & 4.432E+00 & 6.453E+01 & 9.679E+00 & 9.339E+01 & 1.401E+01 & 6.046E+01 & 9.069E+00 & \nodata & \nodata & \nodata & \nodata & \nodata & \nodata & \nodata & \nodata & \nodata & \nodata & \nodata & \nodata & \nodata & \nodata & \nodata & \nodata & \\
J203221.13+402000.9 & 308.088013 & 40.333580 & \nodata & \nodata & \nodata & \nodata & 1.503E-01 & 1.503E-01 & 4.501E+00 & 6.752E-01 & 1.341E+01 & 2.011E+00 & 2.509E+01 & 3.763E+00 & 2.886E+01 & 4.328E+00 & \nodata & \nodata & \nodata & \nodata & \nodata & \nodata & \nodata & \nodata & \nodata & \nodata & \nodata & \nodata & \nodata & \nodata & \nodata & \nodata & \\
J203227.61+401914.5 & 308.115021 & 40.320698 & \nodata & \nodata & 2.947E-01 & 4.421E-02 & 1.975E+00 & 2.962E-01 & 4.053E+00 & 6.079E-01 & 4.727E+00 & 7.090E-01 & 4.491E+00 & 6.736E-01 & 4.919E+00 & 7.378E-01 & \nodata & \nodata & \nodata & \nodata & \nodata & \nodata & \nodata & \nodata & \nodata & \nodata & \nodata & \nodata & \nodata & \nodata & \nodata & \nodata & \\
\enddata
\tablecomments{Table \ref{tab:fluxvalues} is published in its entirety in the machine-readable format.
      A portion is shown here for guidance regarding its form and content.}
\end{splitdeluxetable*}

\begin{table*}\label{28YSOs}
\caption{
           SED-derived parameters for the 28 YSOs in the SMA-observed Region of G79.3+0.3\label{tab:sed_params}}  \centering
  \begin{tabular}{cccccccc}
  \hline
R.A.	&	Dec	&	SMA	&	\Sp &	YSO	&		Mass						& 		Luminosity							& 		Extinction ($A_{v}$)							\\
(degrees)	&	(degrees)	&	ID	&	ID	&	Class	&		($M_{\sun}$)						& 		($L_{\sun}$)							& 		(mag)							\\
\hline																																			
\hline																																			
308.092102	&	40.338074	&	1	&	J203222.10+402017.1	&	1	&	$	5.77	\phantom{0}	\pm	1.73	\phantom{0}	$	& 	$	1543		\pm		470	\phantom{.000}	$	& 	$	62.9	\phantom{0}	\pm		18.6	\phantom{0}	$	\\
308.119202	&	40.328213	&	4	&	J203228.61+401941.6	&	1	&	$	5.17	\phantom{0}	\pm	1.55	\phantom{0}	$	& 	$		\phantom{0} 815	\pm		244	\phantom{.000}	$	& 	$	24.8	\phantom{0}	\pm		7.4	\phantom{00}	$	\\
308.096008	&	40.322975	&	5	&	J203223.04+401922.7	&	2	&	$	4.00	\phantom{0}	\pm	1.20	\phantom{0}	$	& 	$ 37.6		\pm		11.3	\phantom{00}	$	& 	$	7.85		\phantom{0}	\pm	2.35	\phantom{0}		$	\\
\hline																																			
																																			
308.085846	&	40.330597	&	--	&	J203220.60+401950.1	&	1	&	$	4.23	\phantom{0}	\pm	1.27	\phantom{0}	$	& 	$	 268		\pm	80\phantom{000} $	& 	$	91.1	\phantom{0}	\pm		27.3	\phantom{0}	$	\\
308.092682	&	40.332169	&	--	&	J203222.25+401955.8	&	1	&	$	3.29	\phantom{0}	\pm	0.99	\phantom{0}	$	& 	$	29.1		\pm		8.7	\phantom{000}	$	& 	$	40.3	\phantom{0}	\pm		12.1	\phantom{0}	$	\\
308.116089	&	40.328426	&	--	&	J203227.86+401942.3	&	1	&	$	3.25	\phantom{0}	\pm	0.97	\phantom{0}	$	& 	$	34.7	\pm		10.4	\phantom{00}	$	& 	$	30.7	\phantom{0}	\pm	\phantom{0}	9.2	\phantom{0}	$	\\
308.088043	&	40.340446	&	--	&	J203221.13+402025.6	&	1	&	$	3.24	\phantom{0}	\pm	0.97	\phantom{0}	$	& 	$	96.6		\pm		29.0	\phantom{00}	$	& 	$	32.7	\phantom{0}	\pm	\phantom{0}	9.8	\phantom{0}	$	\\
308.083893	&	40.331524	&	--	&	J203220.14+401953.5	&	1	&	$	3.18	\phantom{0}	\pm	0.95	\phantom{0}	$	& 	$	122		\pm	37	\phantom{000}	$	& 	$	54.9	\phantom{0}	\pm		16.5	\phantom{0}	$	\\
308.088013	&	40.333580	&	--	&	J203221.13+402000.9	&	1	&	$	3.15	\phantom{0}	\pm	0.94	\phantom{0}	$	& 	$	67.7	\pm		20.3	\phantom{00}	$	& 	$	81.0	\phantom{0}	\pm		24.3	\phantom{0}	$	\\
308.115021	&	40.320698	&	--	&	J203227.61+401914.5	&	2	&	$	3.07	\phantom{0}	\pm	0.92	\phantom{0}	$	& 	$	15.6		\pm		4.7	\phantom{000}	$	& 	$	33.8	\phantom{0}	\pm		10.1	\phantom{0}	$	\\
308.095093	&	40.328014	&	--	&	J203222.82+401940.9	&	1	&	$	3.00	\phantom{0}	\pm	0.90	\phantom{0}	$	& 	$	32.5	\pm		9.8	\phantom{000}	$	& 	$	59.1	\phantom{0}	\pm		17.7	\phantom{0}	$	\\
308.079529	&	40.336517	&	--	&	J203219.09+402011.5	&	1	&	$	2.41	\phantom{0}	\pm	0.72	\phantom{0}	$	& 	$	9.62	\pm	2.88	\phantom{00}	$	& 	$	100	\phantom{0}	\pm		30	\phantom{00}	$	\\
308.072327	&	40.330471	&	--	&	J203217.36+401949.7	&	2	&	$	2.24	\phantom{0}	\pm	0.67	\phantom{0}	$	& 	$	25.2		\pm		7.6	\phantom{000}	$	& 	$	28.4	\phantom{0}	\pm	\phantom{0}	8.5	\phantom{0}	$	\\
308.095795	&	40.339264	&	--	&	J203222.99+402021.4	&	1	&	$	2.20	\phantom{0}	\pm	0.66	\phantom{0}	$	& 	$	24.0	\pm		7.2	\phantom{000}	$	& 	$	18.4	\phantom{0}	\pm	\phantom{0}	5.5	\phantom{0}	$	\\
308.087952	&	40.313370	&	--	&	J203221.11+401848.1	&	2	&	$	2.09	\phantom{0}	\pm	0.63	\phantom{0}	$	& 	$	4.56\pm		1.37	\phantom{00}	$	& 	$	14.7	\phantom{0}	\pm	\phantom{0}	4.4	\phantom{0}	$	\\
308.074768	&	40.332897	&	--	&	J203217.95+401958.4	&	2	&	$	1.35	\phantom{0}	\pm	0.41	\phantom{0}	$	& 	$	9.41	\pm		2.82	\phantom{00}	$	& 	$	14.0	\phantom{0}	\pm	\phantom{0}	4.2	\phantom{0}	$	\\
308.111969	&	40.319542	&	--	&	J203226.87+401910.4	&	2	&	$	1.35	\phantom{0}	\pm	0.41	\phantom{0}	$	& 	$	9.41	\pm	2.82	\phantom{00}	$	& 	$	16.2	\phantom{0}	\pm	\phantom{0}	4.9	\phantom{0}	$	\\
308.079468	&	40.328423	&	--	&	J203219.07+401942.3	&	1	&	$	1.00	\phantom{0}	\pm	0.30	\phantom{0}	$	& 	$	5.96		\pm		1.79	\phantom{00}	$	& 	$	38.4	\phantom{0}	\pm		11.5	\phantom{0}	$	\\
308.108429	&	40.317829	&	--	&	J203226.02+401904.2	&	2	&	$	0.737		\pm	0.221		$	& 	$	3.44	\pm		1.03	\phantom{00}	$	& 	$	29.0	\phantom{0}	\pm	\phantom{0}	8.7	\phantom{0}	$	\\
308.086060	&	40.314732	&	--	&	J203220.65+401853.0	&	1	&	$	0.577		\pm	0.173		$	& 	$	3.93	\pm		1.18	\phantom{00}	$	& 	$	22.9	\phantom{0}	\pm	\phantom{0}	6.9	\phantom{0}	$	\\
308.084778	&	40.333744	&	--	&	J203220.34+402001.5	&	1	&	$	0.469		\pm	0.141		$	& 	$	0.803		\pm		0.241\phantom{00}		$	& 	$	5.00	\phantom{0}		\pm	\phantom{0}	1.50		$	\\
308.098022	&	40.343136	&	--	&	J203223.52+402035.3	&	2	&	$	0.373		\pm	0.112		$	& 	$	1.29		\pm		0.39	\phantom{00}	$	& 	$	23.5	\phantom{0}	\pm	\phantom{0}	7.0	\phantom{0}	$	\\
308.122589	&	40.321468	&	--	&	J203229.42+401917.3	&	2	&	$	0.214		\pm	0.064		$	& 	$	0.719		\pm		0.216\phantom{00}		$	& 	$	11.7	\phantom{0}	\pm	\phantom{0}	3.5	\phantom{0}	$	\\
308.113464	&	40.323074	&	--	&	J203227.23+401923.1	&	2	&	$	0.173		\pm	0.052		$	& 	$	1.27		\pm		0.38	\phantom{00}	$	& 	$	30.3	\phantom{0}	\pm	\phantom{0}	9.1	\phantom{0}	$	\\
308.078186	&	40.334187	&	--	&	J203218.76+402003.1	&	2	&	$	0.157		\pm	0.047		$	& 	$	0.691		\pm		0.207\phantom{00}		$	& 	$	19.1	\phantom{0}	\pm	\phantom{0}	5.7	\phantom{0}	$	\\
308.077515	&	40.336700	&	--	&	J203218.61+402012.1	&	2	&	$	0.152		\pm	0.046		$	& 	$	0.250		\pm		0.075\phantom{00}		$	& 	$	9.84		\phantom{0}	\pm	\phantom{0}	2.95		$	\\
308.073486	&	40.333916	&	--	&	J203217.64+402002.1	&	2	&	$	0.146		\pm	0.044		$	& 	$	1.23		\pm	0.37	\phantom{00}	$	& 	$	6.20		\phantom{0}	\pm	\phantom{0}	1.86		$	\\
308.084167	&	40.313576	&	--	&	J203220.20+401848.9	&	2	&	$	0.107		\pm	0.032		$	& 	$	0.115		\pm		0.034	\phantom{00}	$	& 	$	6.38	\phantom{0}		\pm	\phantom{0}	1.91		$	\\
\hline																	  
\end{tabular}
\end{table*}

The SEDs of SMA objects 1, 4, and 5 are shown in 
Figure~\ref{fig:sma_seds}. The data points shown are from the near-IR, \Sp, \Her, and SMA. Where the source is not detected in the \Her\ SPIRE bands, upper limits were used, indicated by the downward-pointed triangle symbols. Where the source was not separately resolved in the \Her\ bands (such as SMA sources 1, 2, and 3), an upper limit based on the flux for all sources combined was used in the fitting process.
For SMA source 1, we used the flux values given in the MIPS 24 \micron\ catalog as an upper limit due to multiple sources present.
The models have discontinuities in flux at wavelengths above  
200~\micron\ (e.g., see the left panel of Figure~\ref{fig:sma_seds}) since the signal-to-noise ratio in the models becomes poor 
above 100~\micron\ and the median uncertainty in the model fit is above 
20\% at millimeter wavelengths \citep{robitaille2006}.
Two of the SMA sources each have only one valid fit, likely because the 
input fluxes were too strongly constrained \citep{robitaille2008}. The 
$\chi^{2}$ value for SMA object 1 is large, indicating a poor fit. 
These are a few reasons to only use the SED-derived parameters as first 
estimates, along with further reasons discussed in 
Section~\ref{sec:res_sedparams}.
A selection of the G79.3+0.3 object parameters is given in 
Table~\ref{tab:sed_params} with the uncertainty in each parameter set 
to  $\pm$30\% following \citet{saral2017}.

\begin{figure*}
    \centering
    \includegraphics[width=0.7\textwidth]{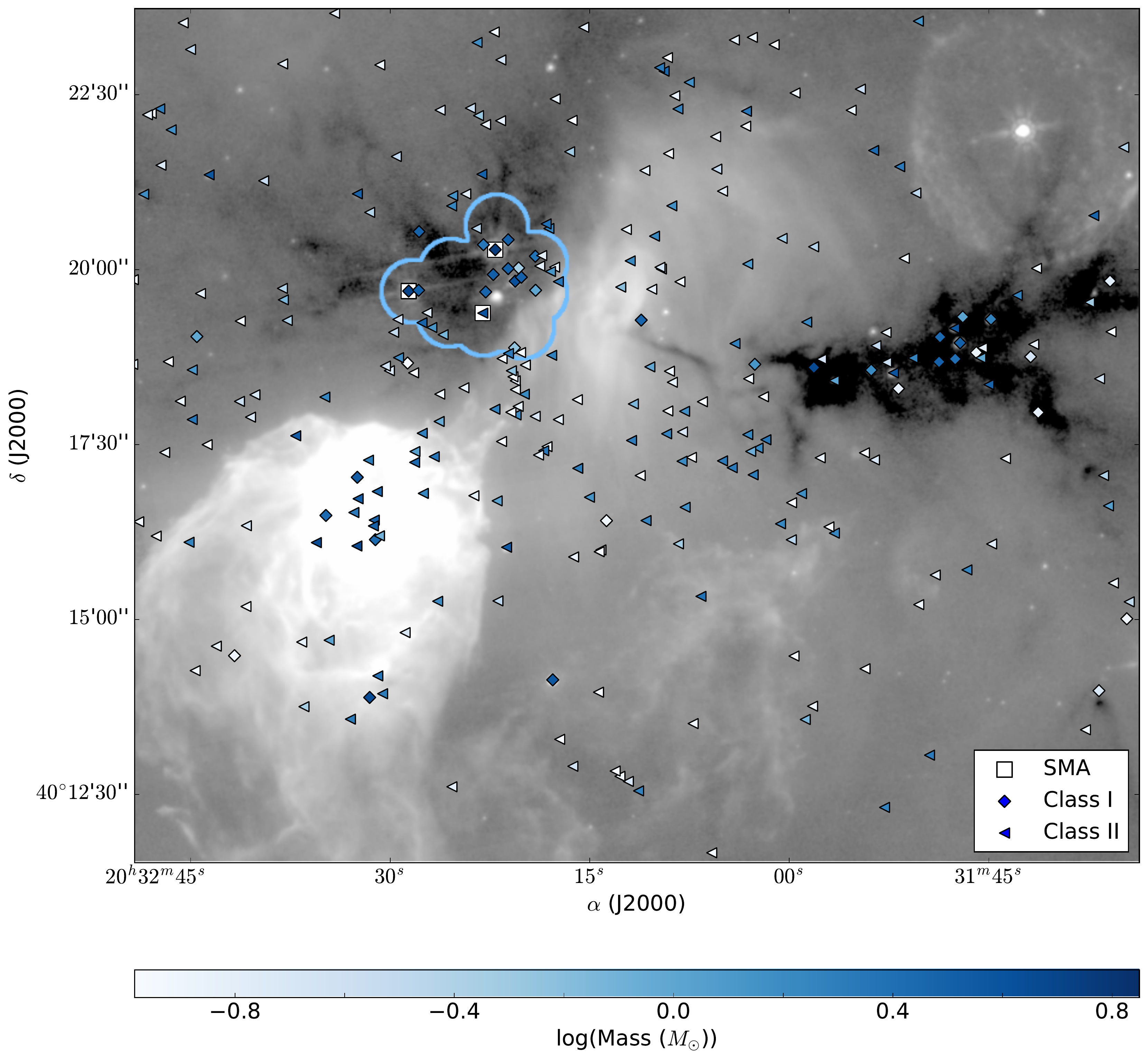}
    \caption{The spatial distribution of YSOs across G79.3+0.3 and 
             its surrounding
             area. The region covered by our SMA data is outlined in 
             blue near the upper left of the image. Class I YSOs are plotted as diamonds,
             Class II YSOs as triangles, and the symbols are shaded according to their mass derived from the SED fitting using the color scale shown below the image, with darker blue being higher mass. Each of the three 
             SMA objects is highlighted with a white square. The 
             background image is the 8~\micron\ \Sp\ image, and the 
             dark black regions are IRDCs.}
    \label{fig:massdist}
\end{figure*}

Figure~\ref{fig:massdist} shows the spatial distribution of YSOs across 
G79.3+0.3 and its surrounding area, with the symbols showing their YSO class and mass. Within G79.3+0.3, the more massive 
objects tend to be near the center of the cloud and located close to 
each other. These more massive objects are almost entirely Class I 
objects, whereas the Class II objects are lower mass and are located on the outskirts 
of the covered area. The distribution is similar in the IRDC region seen on the right side of Figure~\ref{fig:massdist}, which is unsurprising given that the two regions are connected behind the foreground warm dust emission 
\citep{redman2003}.

\subsection{Molecular outflows seen in \textsuperscript{12}CO}
\label{sec:results_12co}

\begin{figure*}
    \includegraphics[width=\textwidth]{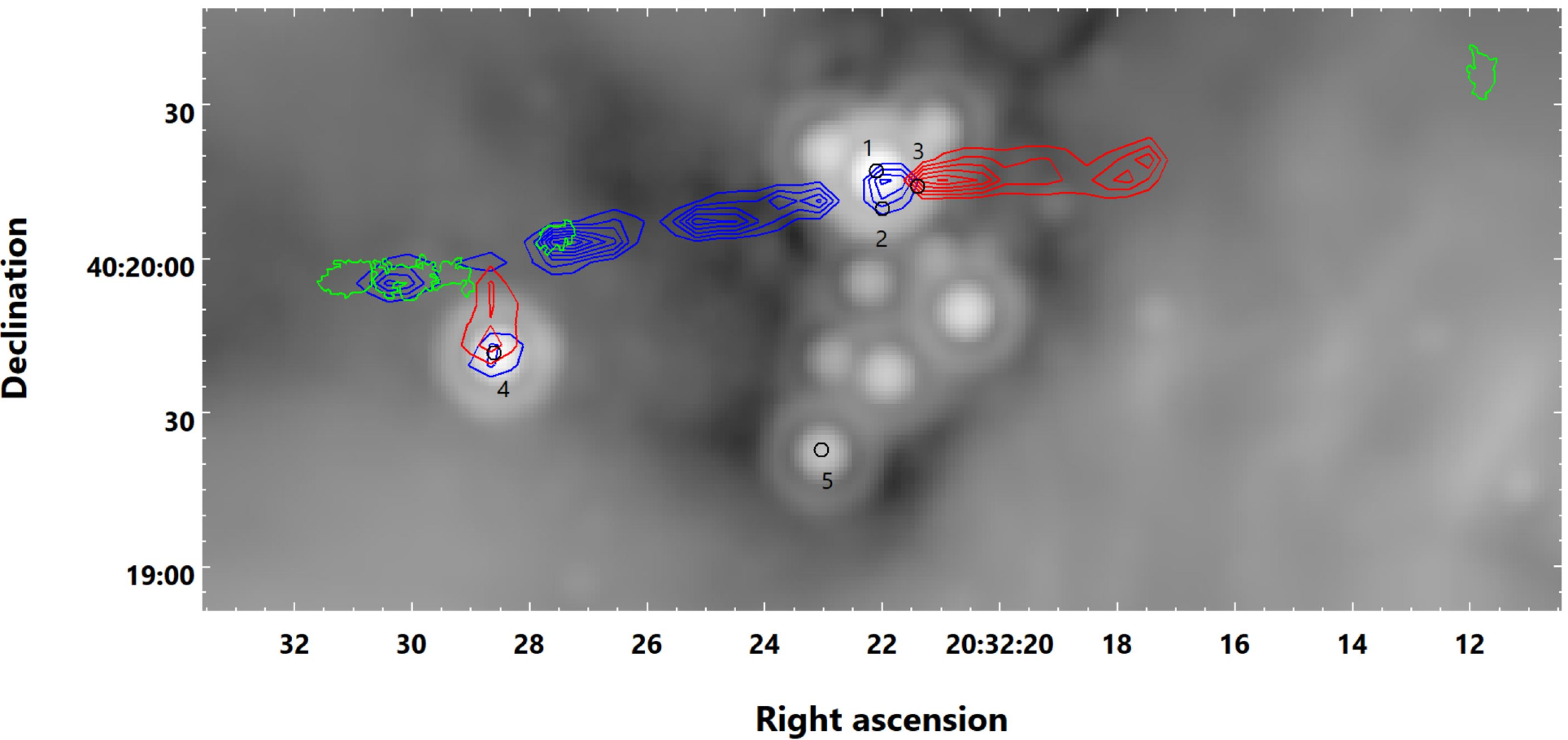}
    \caption{H$_{2}$ and $^{12}$CO tracers of the large outflow from SMA object 3, plotted on the 24~\micron\ \Sp\ image.
     The $^{12}$CO emission is integrated over $7.8$ to $46.8$~km~s$^{-1}$ for the redshifted emission (red contours), and $-31.2$ to 
    $-7.8$~km~s$^{-1}$ for the blueshifted emission (blue contours). 
    The contours are spaced at 7.8~Jy~km~s$^{-1}$ intervals.
    Green contours show the H$_{2}$ emission from \citet{2018ApJS..234....8M}
    as described in Section~\ref{sec:disc_outflow}.
    The black markers labeled 1 through 5 show the locations of the five
    sources in our SMA 1.3~mm continuum image. The origin of the H$_{2}$ 
    and $^{12}$CO is consistent with SMA object 3 being the source of the
    outflow.                }
    \label{fig:h2-12co}
\end{figure*}

To identify molecular outflows in G79.3+0.3, we examined the $^{12}$CO J=2-1 emission in various velocity bins across the region. The dominant CO emission feature in the images is the narrow linear structure at a position angle (PA) of $100.4 \pm 0.4^\circ$. 
This structure is seen in the blueshifted velocities from $-12$ km~s$^{-1}$ up to $-50$ km~s$^{-1}$, and in the redshifted velocities
from $11$ km~s$^{-1}$ up to $45$ km~s$^{-1}$. This high-velocity CO emission traces
a protostellar outflow from a young protostar in the region. 
Following \citet{zhang2005}, we integrate the blueshifted and redshifted CO emission to produce the outflow image shown in Figure~\ref{fig:h2-12co}.
We estimate outflow parameters including its mass,
energy, and momentum using the $^{12}$CO emission.
The mass of the gas can be estimated from the column density, $\bar{N}$(CO). 
Following the formula in \citet{garden1991}, we find the column density from the equation: 
\begin{multline}
\bar{N}(CO) = 4.34\cdot10^{13} \cdot \frac{1}{2}\exp{(5.525/T_{ex})} 
              \cdot \frac{(T_{ex}+0.92)}{\exp{(-11.1/T_{ex})}} \cdot \\
              \int\frac{{T_{B}} \tau dv}{(1-\exp{(-\tau)})}
              .
\end{multline}
The integral is simplified by placing $\bar{N}$ into the $M_{gas}$ 
equation to obtain:
\begin{multline}
M_{gas} = 1.6\cdot10^{-7} \cdot \frac{1}{2}\exp{(5.525/T_{ex})} \cdot 
          \frac{(T_{ex}+0.92)}{\exp{(-11.1/T_{ex})}} \cdot \\
          \frac{\bar{\tau}}{1-\exp{(-\bar{\tau})}} \cdot 
          \left[\frac{\theta}{arcsec}\right]^{2} \int{T_{B}}
          dv M_{\sun}
          .
\end{multline}
This method uses estimates of the mean optical depth $\bar{\tau}$ of CO, and the
excitation temperature $T_{ex}$. We assume that CO is optically thin, and a typical $T_{ex}$ of 20K.

The total outflow mass can be used to derive the outflow momentum 
$P=M(\Delta v$, $\bar{\tau}$)$|\bar{\text{v}}|$ and energy $E=\frac{1}
{2}M(\Delta v, \bar{\tau})\bar{v}^{2}$ for mean 
optical depth $\bar{\tau}$. We can combine these results with the 
dynamical timescale of the outflow t$_{\text{dyn}}$ to find: the rate of 
outflow mass $\dot{M}=M/t_{dyn}$; the mechanical force 
$F=P/t_{dyn}$; and the outflow luminosity $L=E/t_{dyn}$ 
\citep{zhang2005}. 
The dynamical timescale can in turn be found by taking the ratio of the 
maximum extent of the outflow and the maximum speed of the gas. In order to compare with the studies in the literature, we did not correct for the inclination angle of the outflow when estimating outflow parameters.

The $^{12}$CO emission that tracks the major outflow stemming from SMA 
object 3 is shown in Figure~\ref{fig:h2-12co}. The outflow parameters are 
given in Table~\ref{tab:12co_params}. We estimated the spatial extent of the outflow using trigonometry, finding projected lengths 
$L_{R}=0.43$~pc and $L_{B}=0.94$~pc for the red- and blueshifted halves 
of the outflow, respectively. The actual extent of the outflow is likely 
larger, given that it is inclined towards us at some unknown angle.

\section{Discussion}
\label{sec:discussion}

\subsection{SMA Continuum}
The SMA objects 1, 4, 
and 5 can clearly be seen with the infrared instruments featured in Figure~\ref{fig:dr15_panels}. 
It is possible that all three closely-placed SMA objects (1 through 3) 
have significant emission in the \Her\ bands, but were too poorly-resolved to be 
distinguished as individual objects. There is some indication that Source 1 is slightly extended in the direction of Object 3 at 70 and 160~\micron. The dominant emission is still from Source 1 at 160~\micron. However, at the 1.3~mm band Objects 2 and 3 are each more than twice as bright as Object 1, suggesting that it is likely that Objects 2 and 3 are deeply embedded in the cloud, and are at an earlier stage of evolution.

Our SMA continuum map is consistent with a previous observation of 
G79.3+0.3 by \citet{redman2003}, who observed a portion of the IRDC at 3~mm using BIMA and resolved our SMA Objects 1 through 3 (their sources C, B, and A) and seem to have also detected SMA Object 5. Their image shows  
blueshifted HCO$^{+}$(1--0) line emission that is consistent with a 
strong outflow from Object 3, matching the outflow that is traced 
with the $^{12}$CO line emission from the SMA.
The BIMA and SMA observations agree that Object 1 is the most massive 
star in G79.3+0.3. \citet{redman2003} conclude that our SMA Object 1 
will likely evolve into a B star on the main sequence and is too young 
to have disrupted the IRDC and triggered further star formation, but 
that this could happen over the next $10^6$ years.

\citet[Object S37]{motte2007} reported a total mass of 45~$M_{\sun}$ for SMA Objects 1, 2 and 3 based on the MOMBO data at the 1.2~mm band and a resolution of $11''$. The SMA observations have sufficient spatial resolution to distinguish the compact cores. Therefore, the mass estimates in this paper are consistent with that reported in \citet{motte2007}.

SMA Objects 4, 5, and the grouping of 1 through 3 are roughly 
equidistant (the distances are: 1--4, 0.70~pc; 4--5, 0.58~pc; and 1--5, 0.38~pc).
This is similar to the massive IRDC~18223 reported by \citet{beuther2015}, which has twelve cores regularly spaced 
at $0.40\pm0.18$ pc and peak separations varying between 0.19~pc and 0.70~pc, as well as to IRDC G28.34 \citep{zhang2009} with five regularly spaced cores.
The similarity between the inter-core separations in IRDC 18223 and the inter-stellar separations in G79.3+0.3 suggests that our five SMA objects were originally part of the same filamentary structure. 
This can be expanded upon further by considering SMA Objects 1 through 3, 
which are roughly equidistant with a separation of around 0.069 pc,
and so were likely formed from a single massive gas structure that further 
fragmented into the three objects. 
The separations between our SMA objects are comparable to the Jeans length in  G79.3+0.3 of  0.50~pc, assuming values of gas density and kinetic temperature from \citet[Table 3]{carey1998}.
This similarity between inter-star distance and Jeans length is as expected and has been seen in many other clouds \citep[e.g.][]{2006ApJ...636L..45T,2016A&A...592A..21F}.

\subsection{SEDs and Parameters}
\label{sec:res_sedparams}

The SEDs of the three SMA objects are reasonable for a 
first estimate of the object parameters, since the inclusion of SMA data 
gives model SEDs consistent with their previous classifications based on 
\Sp\ data. 

\begin{figure}
\vskip 0.2in
    \centering
    \includegraphics[width=\columnwidth]{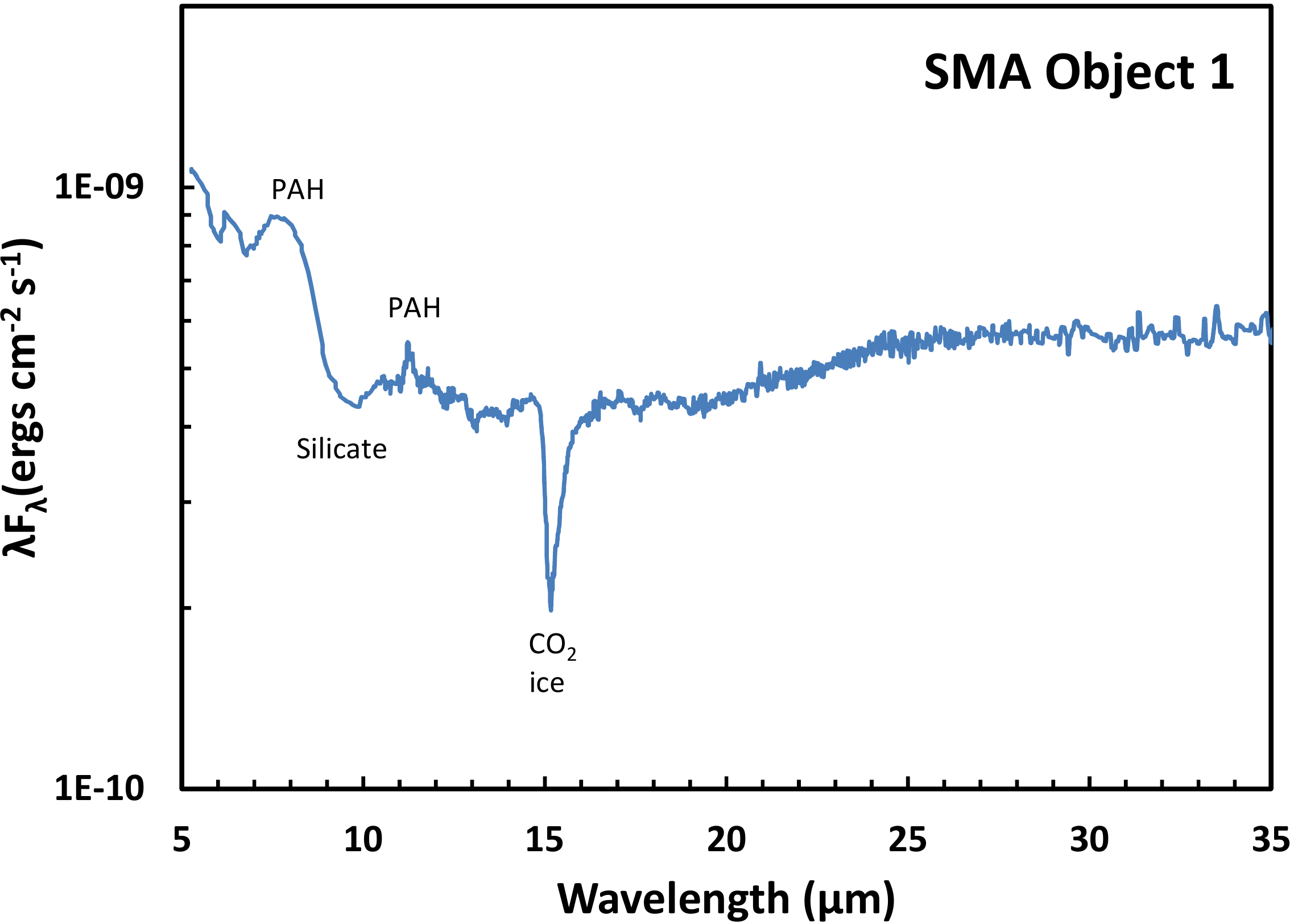}
    \caption{The spectrum of SMA object 1, adapted from 
             \citet[Object CYGXS37]{seguracox2011}. This spectrum was 
             measured using \Sp's InfraRed Spectrograph (IRS). 
             There are emission features due to Polycyclic Aromatic Hydrocarbons (PAHs) at $\sim$ 8 and 11~\micron, and absorption features due to H$_{2}$O ice  
             at 6~\micron,  silicates at $\sim10$~\micron, and CO$_{2}$ ice at $\sim15$~\micron.}
    \label{fig:cygxs37}
\end{figure}

\citet{seguracox2011} reduced a spectrum for SMA Object 1 from data obtained in  \Sp\ project ID 50045 \citep{fazio08}, shown in 
Figure~\ref{fig:cygxs37}. This covers the 5~\micron~$\leq\lambda\leq34$~\micron\ range in much higher detail than our SED in 
Figure~\ref{fig:sma_seds}. 
Both the spectrum and the best-fit SED show silicate absorption at 10~\micron, which is 
consistent with the object being a Class I YSO. Notably, the spectrum 
shows features usually seen in massive YSOs, 
including strong CO$_{2}$ ice absorption at 15.4~\micron\  
and H$_{2}$O ice absorption near 6~\micron\ \citep{an2009}. The 
15.4~\micron\ absorption is caused by mixing of CO$_{2}$ and CH$_{3}$OH,
which are abundant in massive YSOs \citep{an2011}. 
There was some weak CH$_{3}$OH line emission visible in our SMA spectra 
across G79.3+0.3, but the signal-to-noise ratio was too poor to map it across the region. 

It is important to consider the SED-derived parameters as first 
estimates, rather 
than the true physical values of the objects. We are using large 
uncertainties of $\pm$30\%, and the models 
make potentially false 
assumptions of the objects, such as assuming that the YSO envelope is in 
free-fall, that the objects all have solar 
metallicity, and that the disc and envelope gas-to-dust ratios are 100 
\citep{robitaille2006}. 

SMA Objects 1 and 4 are the two most massive and luminous objects in this region of G79.3+0.3, and also have the highest extinction. 
Their high luminosities are consistent with being Class I YSOs that 
will likely become massive stars, as they are already near 
the 8$M_{\sun}$ limit and are still actively accreting mass. 
Objects 1 and 4 also have high extinction values compared with the 
expected values for Cygnus-X of $5\leq\text{A}_{v}\leq10$, which 
indicates that the objects are deeply embedded in the IRDC.
On the other hand, Object 5 has very low extinction and therefore might be in 
the foreground of the dark cloud.

The area surrounding G79.3+0.3 shows many Class II objects of varying mass, few of 
which appear significantly clustered. There is a patch of closely-
placed Class II objects just below our IRDC at approximately 
$20^{\rm{h}}32^{\rm{m}}20^{\rm{s}}$, 
$40\degr18\arcmin00\arcsec$ (J2000.0). These objects could be part of the same 
cloud if it has been obscured by foreground emission, as in the case of 
the connection between G79.3+0.3 and the nearby IRDC seen in 
Figure~\ref{fig:massdist}. For the most part, the Class II objects are 
lower-mass and more spatially distributed than the Class I objects in this region of Cygnus-X.

\subsection{Outflow from SMA Object 3}
\label{sec:disc_outflow}

The outflow parameters we derived have similar magnitudes to the sample of 39 
objects with outflows in \citet{zhang2005}. 

The total mass of the outflow is 0.83~$M_{\sun}$, consistent with the sample of six outflows given in \citet[Table 3]{lee2002}, which 
cover a range 0.01 -- 1.00~$M_{\sun}$. It is also twice as massive as 
any of the six nearby outflows in DR21 found by \citet[Table 4]
{hawley2012}, despite those six outflows being ejected from much more 
massive YSOs. The outflow could be so massive and energetic because 
there is no other outflow detected nearby that could disrupt the flow of 
material or otherwise interfere with the YSO's accretion process.

Other NIR images in the region surrounding G79.3+0.3 have revealed H$_{2}$ emission corresponding to the outflow from SMA Object 3, that lies further out from the star than the $^{12}$CO emission \citep[``Catalogue of Molecular Hydrogen Emission-Line Objects (MHOs) in Outflows from Young Stars" object MHO 3597]{2018ApJS..234....8M,2010A&A...511A..24D}. 
The MHO catalog identifies the source of the outflow as J203222.10+402017.06 (our SMA object 1) \citep{2014AJ....148...11K}, but our SMA observations reveal the true source to be the newly-identified SMA Object 3 (Figure~\ref{fig:h2-12co}). The H$_{2}$ emission extends just beyond the furthest extent of the blueshifted lobe, and about the same distance in the redshifted direction. The SMA field observed ends about halfway to the H$_{2}$ lobe in the redshifted direction, so we cannot tell if the redshifted $^{12}$CO emission extends a similar distance.

\begin{deluxetable}{rrr}
\tablecaption{Estimates of the Parameters for SMA Object 3 Outflow\label{tab:12co_params}}
\tablecolumns{3}
\tablehead{
\colhead{Parameter} & \colhead{Redshifted lobe} & \colhead{Blueshifted lobe}\\
}
 \startdata
  Mass &&               \\
  ($M_{\sun}$)  & 0.37 & 0.46\\
  Momentum            &&\\
  ($M_{\sun}$ km s$^{-1}$)& 10.5 & 14.0\\
  Energy              &&\\
  ($M_{\sun}$ (km s$^{-1}$)$^{2}$) & 163 & 221\\
  Extent              && \\
  (pc) & 0.43 & 0.94\\
  Dynamical timescale &&\\
  (kyr)& 10.5 & 23.0\\
  Mass rate           && \\
  ($10^{-13}\cdot$ $M_{\sun}$ s$^{-1}$) & 11.2 & 6.34\\
  ($10^{-5}\cdot$ $M_{\sun}$ yr$^{-1}$) & 3.53 & 2.00 \\
  Mechanical Force    && \\
  ($10^{-11}\cdot$ $M_{\sun}$ km s$^{-2}$)& 3.17 & 1.93\\
  Luminosity          && \\
  ($10^{-10}\cdot$ $M_{\sun}$(km s$^{-1}$)$^{2}$s$^{-1}$)& 4.91 & 3.05 \\
  ($10^{26}\cdot$ ergs) & 9.76 & 6.06 \\
  \hline
  \enddata
\end{deluxetable}

For all we can find out about the outflow from Object 3, we cannot 
determine much about the central object itself. Object 3 has no previous 
detections in our catalogs, and so we could not fit a model SED and 
estimate the physical properties of this object.

\section{Conclusions}
\label{sec:summary}

We have produced a 1.3~mm continuum image of the IRDC G79.3+0.3 using the SMA. 
The image shows that this region of the IRDC contains
five YSOs in an early stage of formation, one of which has a massive collimated $^{12}$CO 
outflow. The regular spacing of the objects hints at the fragmentation 
scale of the cloud being $\sim0.76$ pc.

We have estimated the properties of the YSOs using model SED fitting. In 
all cases, the slopes of the model SEDs are consistent with the objects' 
 YSO classifications based on color-color and color-magnitude diagrams using the near- and mid-IR data. 
The model SEDs for the three SMA objects 1, 4, and 5 returned masses in the range of 4 -- 6~$M_{\sun}$. This is consistent with 
their being massive YSOs, and in fact they are the most massive and 
luminous of the YSOs we identified in this region of G79.3+0.3. 

We have identified an enormous $^{12}$CO outflow from SMA object 3. The 
outflow extent is at least 0.43~pc and 0.94~pc in the redshifted and 
blueshifted lobes respectively. The presence of this outflow is 
consistent with Object 3 being a protostar, and is also supported by the outflow parameters being of similar magnitude to parameters of 
massive outflows from other studies.
The total mass of the outflow is 0.37 and 0.46~$M_{\sun}$ 
and total momentum is 10.5 and 14.0~$M_{\sun}$~km~s$^{-1}$ 
in the red and blue lobes, respectively.

\section*{Acknowledgements}

The authors wish to recognize and acknowledge the very significant 
cultural role and reverence that the summit of Maunakea has always had 
within the indigenous Hawaiian community.  We are most fortunate to have 
the opportunity to conduct observations from this mountain.

This work is based in part on observations made with the \Sp\ Space 
Telescope, which is operated by the Jet Propulsion Laboratory, 
California Institute of Technology under a contract with NASA. Support 
for this work was provided by NASA through an award issued by 
JPL/Caltech.

\Her\ is an ESA space observatory with science instruments provided 
by European-led Principal Investigator consortia and with important 
participation from NASA.

This publication makes use of data products from the Two Micron All Sky 
Survey, which is a joint project of the University of Massachusetts and 
the Infrared Processing and Analysis Center/California Institute of 
Technology, funded by the National Aeronautics and Space Administration 
and the National Science Foundation.

This work is based in part on data obtained as part of the UKIRT 
Infrared Deep Sky Survey.

The authors wish to thank S.V. Makin for providing the data on H$_{2}$ 
emission in the G79.3+0.3 region. 
The authors thank the anonymous referee for their useful comments that have improved our paper.


\bibliographystyle{aasjournal}
\bibliography{ms}


\software{MIR (\url{https://github.com/qi-molecules/sma-mir}), CASA \citep{2007ASPC..376..127M}, MIRIAD \citep{1995ASPC...77..433S}, SEDFitter \citep[v1.0][]{robitaille2007}}

%


\end{document}